# Generalized analytical relations to describe global optical systems with a plenoptic camera

MARC BRUNEL*, PIERRE SCHLEUNIGER, ADRIAN MARSZALEK, VALENTIN MUSET, RAMANA KUKKARASI, SEBASTIEN COETMELLEC, JEAN-BERNARD BLAISOT

[1]UMR CNRS 6614 CORIA, Avenue de l'Université, BP 12, 76801 Saint-Etienne du Rouvray cedex, France
*Corresponding author: brunel@coria.fr

**The optical transfer matrix formalism is used to describe global set-ups incorporating a plenoptic camera. Analytical relations that give the effective resolution, depth of field, disparity and optimum patch size for image reconstruction are established versus the optical parameters of any global arrangement. The potentiality of this formulation is illustrated analyzing experimental results obtained in astigmatic cylindrical imaging conditions.**

## 1. INTRODUCTION

Light field imaging has emerged as a powerful measurement and analysis technique in modern metrology. Instead of recording only the intensity of light at each pixel, a light field camera also measures the direction of incoming rays. This additional angular information enables 3D reconstructions, refocusing after capture, and robust depth estimation, from a single exposure. These capabilities open opportunities for precise, non-contact three dimensional measurements [1] [2] [3] [4].

Light field imaging is particularly compelling for the metrology of sprays, a domain where traditional measurement techniques often struggle with high speeds, dense droplet fields, occlusions, and complex light–matter interactions. The reconstruction of 3D information from a single snapshot represents an invaluable advantage when characterizing rapidly evolving spray structures [5] [6] [7] [8]. In spray metrology, accurate measurements of droplet size, velocity, number density, and spatial distribution are essential for applications ranging from combustion and propulsion to agricultural spraying and pharmaceutical aerosols. Conventional imaging approaches typically require multi-camera setups with precise alignments and synchronization. In contrast, light field imaging enables computational refocusing and volumetric reconstruction without multiple cameras. In order to analyze real flows, the optical system often becomes more complex than a standalone plenoptic camera. Achieving sufficient spatial resolution may require custom imaging configurations. The set-up environment (flows in pipes, in combustion or high pressure chambers with limited optical access and imposed working distance) imposes additional constraints. Integrating these with a light-field sensor frequently calls for additional relay lenses and optical windows. These additions transform the system into an optical platform in which the plenoptic sensor is only one part of a larger metrological instrument. While the plenoptic camera is central to capturing the light field, practical measurements require a custom-engineered optical architecture. In such cases, it becomes difficult to optimize the plenoptic set-up in the global optical arrangement. The existence of general analytical relations that link the main properties of the plenoptic function to the complex optical arrangement would be particularly valuable for system design, data analysis and interpretation. In this study, we use the optical transfer matrix formalism to establish general analytical relations that give the effective resolution, depth of field, disparity and optimum patch size for image reconstruction versus the parameters of any optical arrangement [3]. The use of these relations is illustrated analyzing experimental results obtained in complex configurations. Section 2 details the procedure developed and the relations obtained. Section 3 presents some applications, in particular the case of 3D-reconstruction through a cylindrical astigmatic pipe.

## 2. OPTICAL TRANSFER MATRIX FORMALISM FOR OPTICAL SYSTEMS INVOLVING PLENOPTIC CAMERAS

### A. Configuration of a plenoptic camera and notations

This first section defines our notations (that are those of reference [9]) and recalls some basic formula that are often used to describe plenoptic camera. Let us consider the system composed of a main lens and a microlens as depicted on figure 1. Both lenses are assumed to be thin lenses. The initial object $X_i$ is at algebraic distance $a_L$ from the main lens. Its image $X_0$ by the main lens is at algebraic distance $b_L$ (Note that index $L$ will refer to the main lens). For the microlens, this intermediate $X_0$ acts as a virtual object at algebraic distance $a$ from the microlens, while the final focused image $X_1$ is at algebraic distance $b$. In a plenoptic configuration, the sensor is not necessarily at distance $b$ from the microlens (depending on the depth position of the initial object). The algebraic distance between the microlens and the sensor will be noted $B$ (the distance from the main lens to the sensor is $B_L$ ). The algebraic positions $a$ and b are linked by the conjugation law $\frac{1}{b}=\frac{1}{f}+\frac{1}{a}$ where $f$ is the image focus length of the microlens.

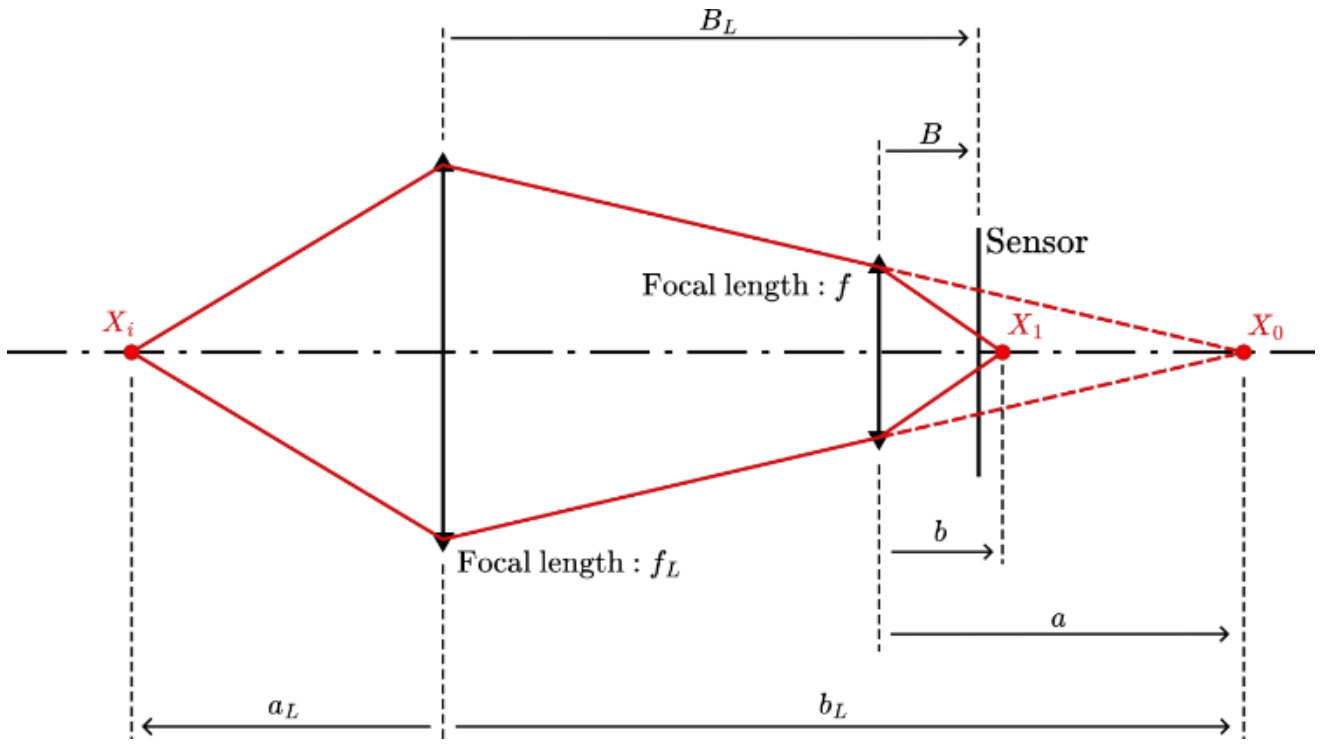

Fig 1 : Schematic configuration of a plenoptic camera considering only one microlens (focus length f).

We note p the pixel size of the sensor, $s_\lambda = 1.22\,\lambda\,N$ the Airy diameter on the sensor, where $N = B/D$, with $D$ the diameter of the microlens. We further define $s_0 = max[p, s_\lambda]$ and $s$ the geometric diameter of the defocused image on the sensor of the on-axis object point $X_i$. Geometrically, we have [9]:

$$s = \left|D\left(\frac{B}{b}-1\right)\right| = \left|D\left(\frac{B}{f}+\frac{B}{a}-1\right)\right| \qquad (1)$$

This relation allows to express the depth of focus (see appendix A). The absolute value in this relation enables to describe the two possible configurations, i.e. when the final focused image $X_1$ is before or after the sensor ($b < B$ and $b > B$). In the case of the plenoptic configuration, the virtual depth is defined by $v = a/B$. If we consider the plenoptic camera in the 1D case as in figure 1, the virtual depth gives the number of images of a point generated by the microlenses. For example, a point at virtual depth $|v| = 3$ is imaged by three adjacent microlenses.

## B. Optical transfer matrix formalism

Appendix B introduces briefly the optical transfer matrix formalism. Let us illustrate the conjugation law through an optical system using this optical transfer matrix formalism. Consider the system of figure 2, where $\boldsymbol{T_{int}} = \begin{bmatrix} T_{11} & T_{12} \\ T_{21} & T_{22} \end{bmatrix}$ is the matrix of the central part of the system. The whole matrix from the object to the image is:

$$\begin{bmatrix} T_A & T_B \\ T_c & T_D \end{bmatrix} = \begin{bmatrix} 1 & z_f/n_f \\ 0 & 1 \end{bmatrix} \begin{bmatrix} T_{11} & T_{12} \\ T_{21} & T_{22} \end{bmatrix} \begin{bmatrix} 1 & z_i/n_i \\ 0 & 1 \end{bmatrix} \qquad (2)$$

If we know the position of the focused image, i.e. $z_f$, the position of the focused object $z_i$ will be given by the condition $T_B = 0$, which gives :

$$z_i = -\frac{T_{12}+T_{22}\frac{z_f}{n_f}}{\frac{T_{11}}{n_i}+\frac{T_{21}\,z_f}{n_i\,n_f}} \qquad (3)$$

Reciprocally, the position of the focused image versus the position of the object is given by:

$$z_f = -\frac{T_{12}+T_{11}\frac{z_i}{n_i}}{\frac{T_{22}}{n_f}+\frac{T_{21}\,z_i}{n_i\,n_f}} \qquad (4)$$

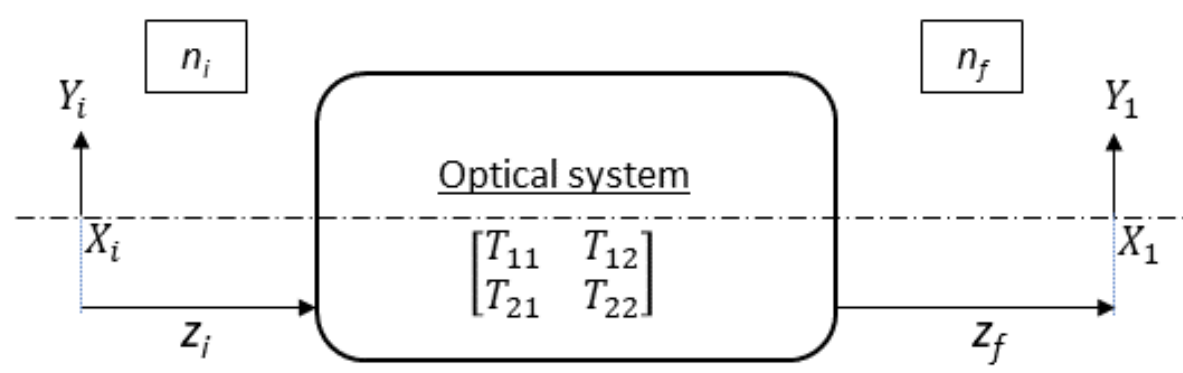

Fig. 2: Conjugation law with optical transfer matrix formalism

This optical transfer matrix formalism can be used in the case of a system involving a plenoptic camera to address rapidly different points:

- Predict the object's domain $[a_L^+; a_L^-]$ where the geometrical image of a point object given by a sole microlens does not exceed the maximum of pixel size and Airy spot size (i.e. $s \leq s_0 = max[p, s_\lambda]$).
- Predict the effective resolution versus the longitudinal position of the object. It corresponds to the ratio between the resolution of the plenoptic camera over the maximum resolution achievable with the sensor.
- Predict the size of the defocused image of a point object versus its longitudinal position.
- Predict the disparity versus the longitudinal position of the object.
- Predict the size of the patch to be used versus the longitudinal position of the object.
- Make these predictions in the case of any complex imaging system.
- In particular, perform such predictions in the case of cylindrical imaging systems, where the optical transfer matrices for the two transverse axes of the system are different.

Next section will address these different points.

**C. Analytical relations for the description of a general system incorporating a plenoptic camera**

*C.1. System under consideration*

Let us now consider the case of a plenoptic camera that is the last part of a global imaging set-up as depicted in figure 3. The index of the output medium is $n_f = 1$ in this case.

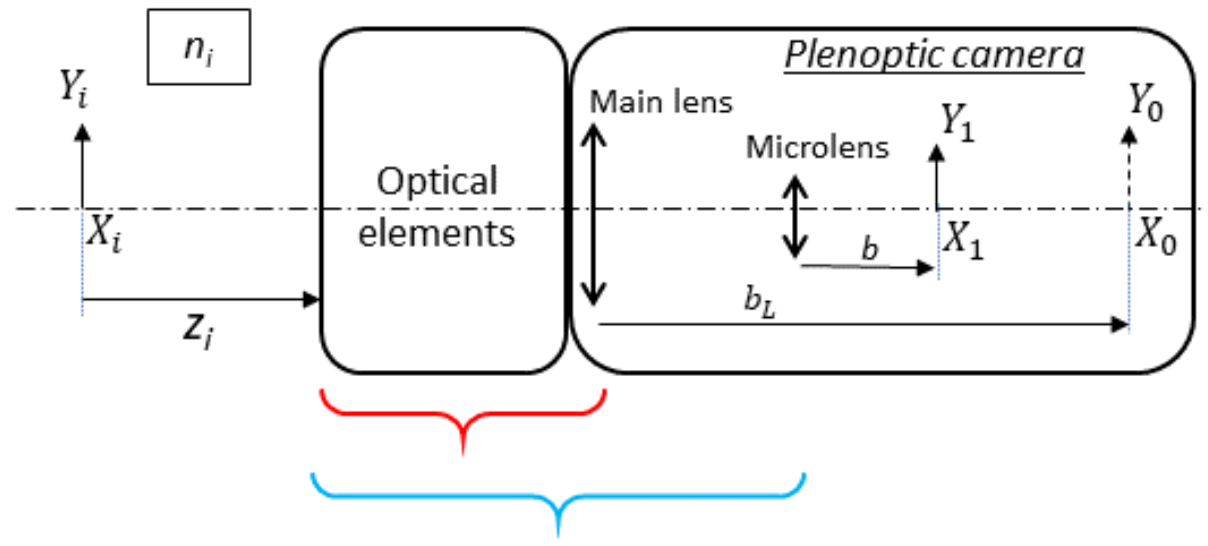

Fig. 3: General imaging set-up with a plenoptic camera

Depending on the expected result, two descriptions can be developed:

(i) We can either express the position of the object $X_i$ versus the position of the intermediate image $X_0$, i.e. parameter $b_L$ on figures 1 and 3. The matrix of the central part $\boldsymbol{T_{int}}$ is then simply the matrix corresponding to the association : optical elements + main lens (focus length $f_L$) (red bracket on figure 3). This description will be called "intermediate description" in what follows.

(ii) Or we can express the position of the object $X_i$ versus the position of the final image $X_1$, i.e. parameter $b$ in figures 1 and 3. The matrix of the central part $\boldsymbol{T_{int}}$ corresponds then to the following components: optical elements + main lens (focus length $f_L$) + free space over distance $(B_L - B)$ + microlens (focus length $f$) (blue bracket in figure 3). Parameters $B_L$ and $B$ are defined in figure 1. $(B_L - B)$ is the distance between the main lens and the microlens. This description will be called "final image description" in what follows.

*C.2. Depth of field*

Let us consider the system of figure 1. The depth of focus is limited by the two extrem positions $a_L^+$ and $a_L^-$ of the object $X_i$ (conjugated with the two positions $a^+$ and $a^-$ of the intermediate image $X_0$, and the two positions $b^+$ and $b^-$ of the final image $X_1$ respectively). They correspond to the limit cases where the size $s$ of the defocused image on the CCD sensor (see relation (1)) is exactly equal to parameter $s_0$. Both positions are illustrated in figures A1a and A1b of appendix A.

Let us see how to determine positions $a_L^+$ and $a_L^-$ with the transfer matrix formalism using both intermediate and final image descriptions. when there are no additional "optical elements", i.e. when the imaging system reduces to the sole plenoptic camera. In the case of the intermediate description (red bracket on the figure 3), the matrix of the intermediate part is:

$$\boldsymbol{T_{int}} = \begin{bmatrix} 1 & 0 \\ -\frac{1}{f_L} & 1 \end{bmatrix} \qquad (5)$$

According to figures 1 and 3, we can write in this case $z_i = -a_L$ (the object is assumed surrounded by air: $n_i = 1$). $a_L$ is algebraic and negative for a real object $X_i$ as illustrated in figure 1. And we have $z_f = b_L$ ($b_L$ is positive on both figures 1 and 3). According to relation (3), the position of the initial object is then given by:

$$z_i = -a_L = -\frac{b_L}{1-\frac{b_L}{f_L}} = -\frac{f_L\, b_L}{f_L - b_L} \qquad (6)$$

This is actually the conjugation law by the main lens. As $b_L = B_L + a - B$, we obtain the expression of $a_L$ versus parameter $a$. It enables to obtain the depth of focus of the system. The determination of two extreme positions, $a_L^+$ and $a_L^-$, from the positions of the intermediate images $a^+$ and $a^-$ respectively (see relations (A3) and (A4) in appendix A), is obtained as well. Relation (6) applies to a simple system consisting of the single main lens. If additional optical elements are present (see Fig. 3), this relation can now be generalized to any imaging system through relation (3) that depends on the optical matrix coefficients of any system with additional optical elements.

For the final image description (blue bracket in figure 3), the matrix of the intermediate part $\boldsymbol{T_{int}}$ (main lens + free space over distance $(B_L - B)$ + microlens) is given by:

$$\boldsymbol{T_{int}} = \begin{bmatrix} 1 - \frac{B_L - B}{f_L} & B_L - B \\ -\frac{1}{f} - \frac{1}{f_L} + \frac{B_L - B}{f\, f_L} & 1 - \frac{B_L - B}{f} \end{bmatrix} \qquad (7)$$

In this case, the position $z_i$ of the initial object (or the algebraic position $a_L$) is given by:

$$z_i = -a_L = -\frac{T_{12} + T_{22}\, b}{T_{11} + T_{21}\, b} \qquad (8)$$

As previously, it enables to obtain the positions of $a_L^+$ and $a_L^-$ associated to the positions $b^+$ and $b^-$ of the final images respectively (relations (A1) and (A2) given in appendix A). It is easily verified that the expressions of $a_L^+$ and $a_L^-$ obtained are rigorously the same with both methods (intermediate and final image descriptions). Replacing $b$ by $B$ in relation (8), it is further possible to determine the position of the object's plane that gives a focused image in the plane of the sensor :

$$z_{i,0} = -a_{L,0} = -\frac{T_{12} + T_{22}\, B}{T_{11} + T_{21}\, B} \qquad (9)$$

*C.3. Effective resolution*

The effective resolution ratio for a single microlens can be defined by $\epsilon_L = \frac{p}{\max[s_0, s]}$. The effective resolution ratio of a plenoptic camera in the 1D case is then given by [9]:

$$\epsilon_p = \frac{1}{|v|}\,\epsilon_L = \frac{1}{|v|}\,\frac{p}{\max\left[s_0, \left|D\left(\frac{B}{f}+\frac{1}{v}-1\right)\right|\right]} \quad (10)$$

With the generalized formalism, the formula to obtain the effective resolution is still given by relation (10) with the virtual depth $v$ given by :

$$v = \frac{a}{B} = \frac{b_L-(B_L-B)}{B} = \frac{-\frac{T_{12}+T_{11}\,\frac{z_i}{n_i}}{T_{22}+\,T_{21}\,\frac{z_i}{n_i}}-(B_L-B)}{B} \quad (11)$$

This last relation is deduced from relation (4) with $z_f = b_L$. In this case, as parameter $a$ is needed to define parameter $v$, we use the intermediate description, where the matrix of the central part $\boldsymbol{T_{int}}$ is the matrix corresponding to the association: optical elements + main lens (red bracket in figure 3).

*C.4. Optimum patch size for reconstruction and disparity*

One can reconstruct an image by sampling a patch of pixels on each microimage and tile them accordingly to their arrangement within the raw image [3]. Each object has an optimum patch size M directly linked to its position along the optical axis. When reconstructing an image with a plenoptic 2.0 camera (focused plenoptic camera), the optimal size of the patch is given by $M = \mu\,\left|\frac{b}{a}\right|$ in the case of proper focusing, where $\mu$ is the distance between microlenses (pitch of the microlens array); $b$ and $a$ are the parameters already used. They represent the positions of the final focused image $X_1$ and the intermediate image $X_0$ from the plane of the microlenses respectively. This relation is illustrated in figure 4.

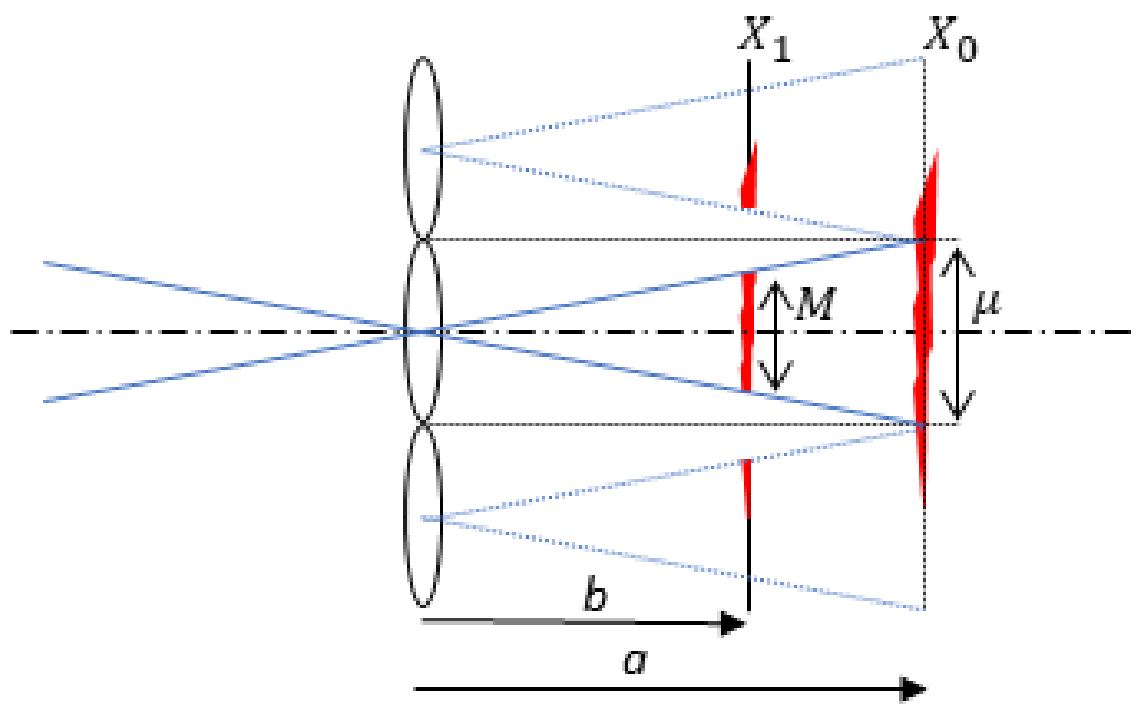

Figure 4: Image capture with optimum patch size.

For a focused image on the CCD sensor, $b = B$ (see figure 1). The determination of the patch size requires the knowledge of parameter $a$ . We will use intermediate description, i.e. the definition of an intermediate transfer matrix $\boldsymbol{T_{int}}$ that includes the optical elements and the main lens. According to relation (4), we can write:

$$b_L = -\frac{T_{12}+T_{11}\,\frac{z_i}{n_i}}{T_{22}+T_{21}\frac{z_i}{n_i}} \quad (12)$$

As parameter $a$ is then given by $a = b_L - B_L + B$ , we obtain for the optimum patch size:

$$M = \frac{\mu\,B}{\left|-\frac{T_{12}+T_{11}\,\frac{z_i}{n_i}}{T_{22}+T_{21}\frac{z_i}{n_i}}-B_L+B\right|} \quad (13)$$

The patch size in light field imaging is directly correlated to the disparity that is linked to the depth of the object. It is illustrated in figure 5. For an intermediate image $X_0$ given by the main objective, two neighbored microlenses of optical centers $O_1$ and $O_2$ will provide final images with disparity, as in conventional stereoscopic set-up. The disparity is given by: $d = x_1 - x_2 = \frac{\mu\,B}{a}$ (with $x_2 < 0$ in figure 5). $\mu$ is the MLA pitch while $a = b_L - B_L + B$ according to figure 1. This parameter measured from adjacent micro-images is essential to deduce the depth of an object.

As previously seen, the virtual distance $a$ is obtained using the matrix transfer formalism, which leads to the general relation, similar to the expression of the patch size:

$$d = \frac{\mu\,B}{-\frac{T_{12}+T_{11}\,\frac{z_i}{n_i}}{T_{22}+\,T_{21}\,\frac{z_i}{n_i}}-(B_L-B)} \quad (14)$$

Reciprocally, the depth position of the object is linked to the disparity through relation:

$$\frac{z_i}{n_i} = -\,\frac{T_{12}+T_{22}\,\left(\frac{\mu\,B}{d}+B_L-B\right)}{T_{11}+T_{21}\,\left(\frac{\mu\,B}{d}+B_L-B\right)} \quad (15)$$

Note that when the disparity $d$ equals 0, we obtain: $z_i = -\,n_i\,\frac{T_{22}}{T_{21}}$. This position is nothing else than the position of the object's focus point of the optical system in red bracket of figure 3 (optical elements + main lens). The virtual distance $a$ is then infinite, corresponding to the situation where point $X_0$ is at infinity in figure 5 (there is no disparity).

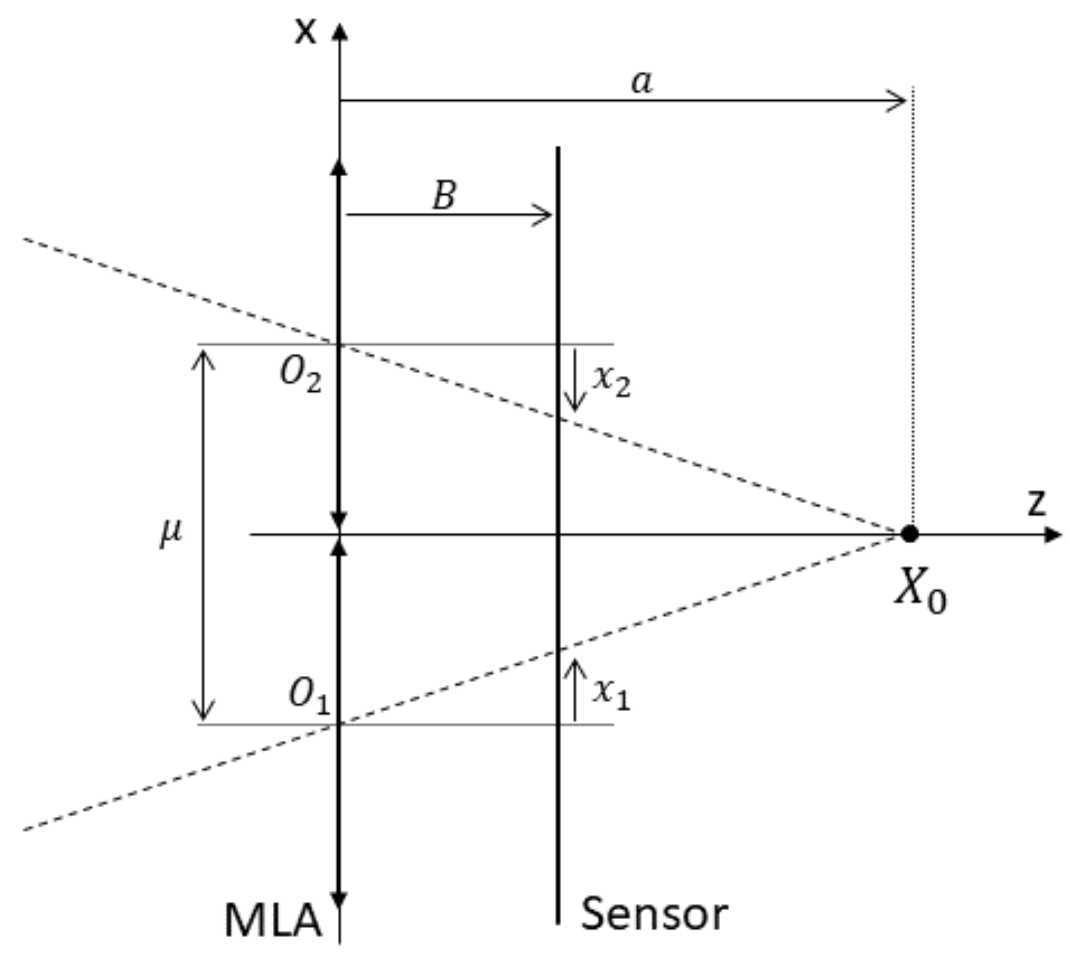

Figure 5: conventional stereoscopic set-up with two sensors and their two optical centers.

*C.5. Defocused image's diameter*

Another possibility with this matrix transfer formalism is to estimate the size $s$ of the geometrical defocused image given by a microlens of the system in the plane of the sensor (for a point object at any depth). In this case we consider the optical matrix of the whole system : from the object's plane to the sensor's plane. In this case:

$$\begin{bmatrix} T_A & T_B \\ T_c & T_D \end{bmatrix} = \begin{bmatrix} 1 & B \\ 0 & 1 \end{bmatrix} \begin{bmatrix} 1 & 0 \\ -\frac{1}{f} & 1 \end{bmatrix} \begin{bmatrix} 1 & B_L - B \\ 0 & 1 \end{bmatrix} \begin{bmatrix} 1 & 0 \\ -\frac{1}{f_L} & 1 \end{bmatrix} \begin{bmatrix} 1 & a_L \\ 0 & 1 \end{bmatrix} \quad (16)$$

Considering an on-axis point object (see figure 6 for illustration), the limit output ray satisfies:

$$y_{2,lim} = \frac{T_B}{T_D}\, \theta_{2,lim} \quad (17)$$

with $\theta_{2,lim} = \frac{y_{2,lim} - \mu/2}{B}$ for the ray exiting from the border of the aperture (in paraxial conditions where $tan(\theta_2) \approx \theta_2$).. Both $y_{2,lim}$ and $\theta_{2,lim}$ are negative on figure 6. These relations lead directly to a geometrical defocused image diameter $s$ in the plane of the sensor:

$$s = 2\,|y_{2,lim}| = \left|\frac{-T_B\,\mu}{T_D\,B - T_B}\right| \quad (18)$$

This relation assumes that the exit pupil is the microlens aperture, which is valid as long as the virtual depth is higher than 1 with a plenoptic 2.0 camera. If the limit angle is approximated by the output aperture angle $\left(-\frac{\mu}{2\,B}\right)$, this relation simplifies to $s \approx \left|\frac{T_B}{T_D}\right| \frac{\mu}{B}$.

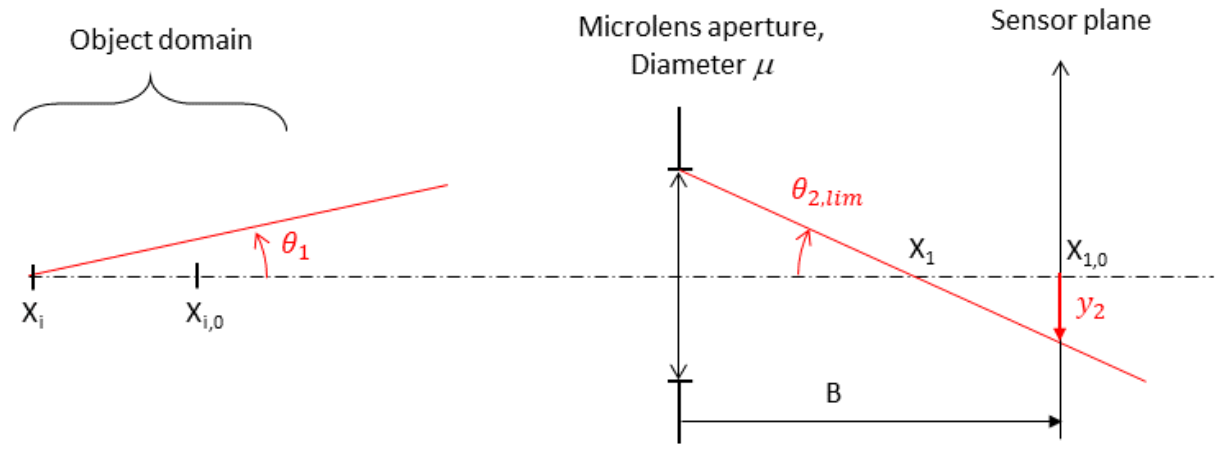

Figure 6: Determination of the size of a defocused image.

## 3. APPLICATIONS

### A. Main objective described by its principal points

The depth of field (DOF) is a very sensitive parameter in the case of spray imaging with a plenoptic 2.0 camera. A thin lens model does not describe accurately the real objective used as main lens, which tends to give erroneous estimations of the DOF. The first application we consider is thus the case of a main lens that cannot be considered as a thin lens. Let us for example suppose that the main lens can be represented by its principal points $H_L$ and $H'_L$, with an unchanged focus length $f_L$. The optical transfer matrix of this element between the planes of $H_L$ and $H'_L$ is given by: $T_{H_L H'_L} = \begin{bmatrix} 1 & 0 \\ -\frac{1}{f_L} & 1 \end{bmatrix}$. But in this case the points $H_L$ and $H'_L$ do not coincide with the center $O_L$ of the lens when it was assumed thin. Considering the calculations of previous sections in the case of this thick system model, the matrix of the thin lens should be replaced by the product: $\begin{bmatrix} 1 & \overline{H'_L O_L} \\ 0 & 1 \end{bmatrix} \begin{bmatrix} 1 & 0 \\ -\frac{1}{f_L} & 1 \end{bmatrix} \begin{bmatrix} 1 & \overline{O_L H_L} \\ 0 & 1 \end{bmatrix}$.

To illustrate this, figure 6(a) shows the effective resolution ratio of the plenoptic camera versus the position of the initial object (parameter $a_L = \overline{O_L X_\iota}$) for two cases: a thin lens in blue, and a thick lens model in red. The main lens is placed at center $O_L$ for the thin-lens hypothesis, and is represented by its principal points $H_L$ and $H'_L$ for the thick-lens hypothesis, with $\overline{O_L H_L} = -4\ mm$, and $\overline{H'_L O_L} = 0mm$. Other parameters are $B_L = 164mm$, $B = 1.7mm$, $f_L = 100mm$, $f = 1.55\ B$, $\mu = 200\ \mu m$, and a pixel's size of the sensor $p = 10\mu m$. $s_\lambda$ is estimated considering a wavelength of $580nm$. We observe a shift of 4mm between the red and blue curves. The formalism developed enables to describe any thick system from the knowledge of its principal points. The limit positions $a_L^+$ and $a_L^-$ obtained using relations (8), (A1) and (A2) are also reported as dashed-dotted lines. They correspond perfectly to the positions of slope breaks of the resolution curves plotted using relations (10-11).

To complete this, figure 6(b) shows the optimum patch size to be chosen versus the position of the initial object (using relation (13)). The blue curve considers a thin lens, while the red curve considers the thick system model. The curves are plain in the depth-of-focus domains (between the limit positions $a_L^+$ and $a_L^-$).

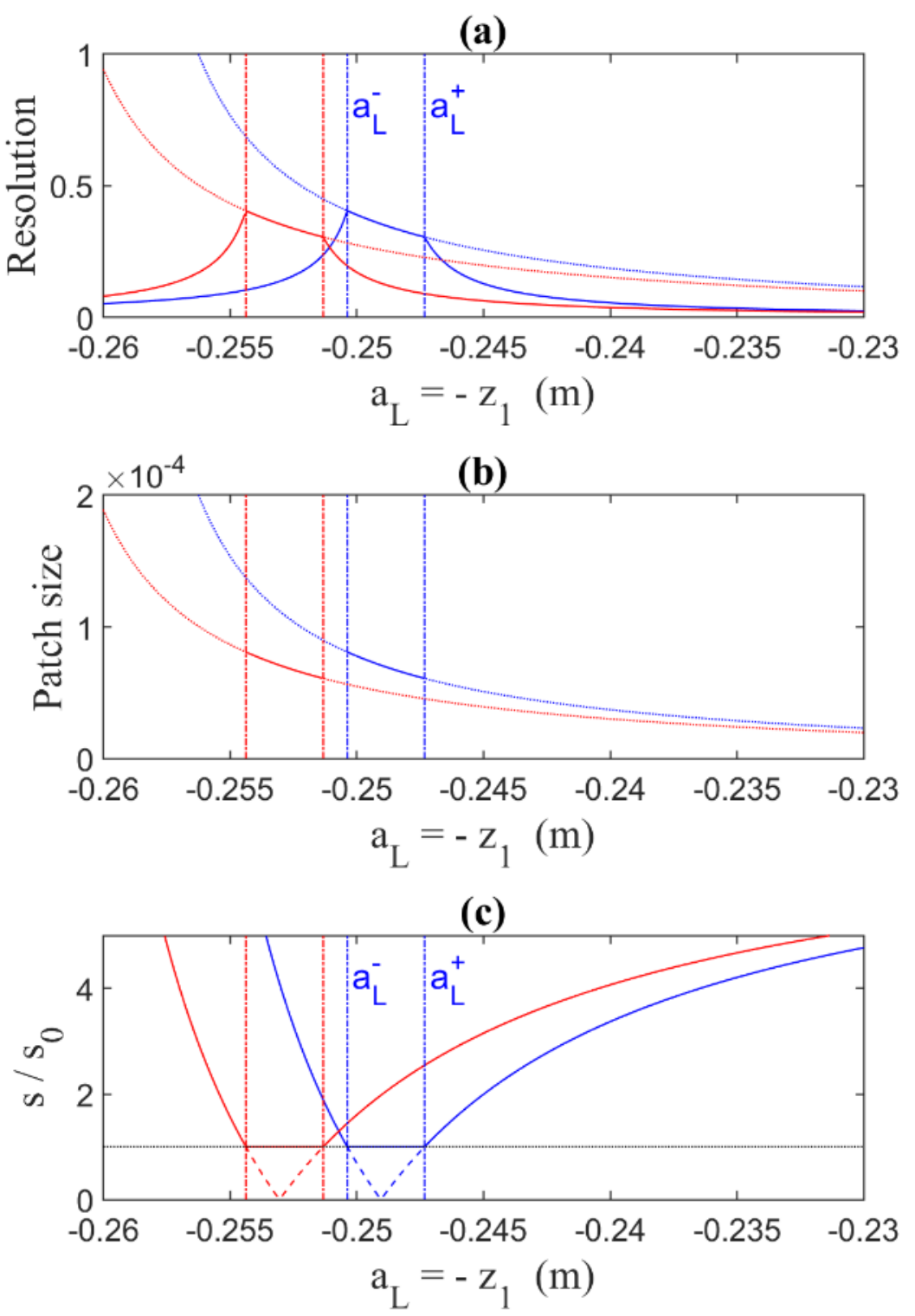

Figure 6: Effective resolution (a), Optimal patch size (y-axis in meters) (b) and image's size of a point object (c) versus the initial object's position (thin lens in blue, system described by its principal points in red).

Finally, figure 6(c) shows the diameter $s$ of the geometrical defocused image of a point object on the sensor versus the longitudinal position of the object. It is the size of the image given by one microlens. It is normalized with respect with the smallest possible size $s_0 = max[p, s_\lambda]$ (p the pixel size and $s_\lambda$ the Airy diameter on the sensor). The blue curves are obtained assimilating the main lens to a thin lens. The dashed line represents the diameter $s$ obtained using relation (18). The plain line represents $max[s, s_0]$. Both curves superpose when $max[s, s_0] = s$. The red curves are obtained with the thick system model, representing the main lens by its principal points. The parameters used are those of previous figures 6 (a) and (b). As in figures (a) and (b), we observe the same shift between the red and blue curves linked to a thin or thick lens approximation. The limit positions $a_L^+$ and $a_L^-$ are also reported. They correspond perfectly to the positions where $s$ reaches the smallest size $s_0$.

### B. Imaging through a cylindrical set-up. Comparison with experimental results

3D-charaterizations in the domain of energy require often in situ measurements which impose severe constraints (flows in pipes or in high pressure chambers with limited optical access for instance). Additional optical elements as windows or image transfer systems are necessary. The presence of cylindrical geometries is one of these configurations that greatly complicates data analysis. It is then necessary to develop accurate modelling and/or calibration procedures to perform quantitative analyses, as could be done with digital holography [11] [12], interferometric particle imaging [13] [14] or with light field imaging [10]. This last recent study presented the characterization of bubbles in a cylindrical pipe using a multifocus 2.0 plenoptic camera [10]. Combining three types of microlenses with different focus lengths, such cameras offer an extended depth of field [9]. The cylindrical set-up that was used is presented in figure 7. The central element of the set-up is a cylindrical glass tank of diameter 20 cm filled with demineralized water. An injector is centered at the bottom of the tank and generates bubbles that rise up to the surface. The flow is illuminated with a diffuse backlight illumination set-up. A Raytrix R29 plenoptic camera is placed after the cylinder ; It is centered so that the optical axis of the camera and the injector's axis intersect. Two configurations were used, one with the camera close to the bubbly flow giving the highest magnification (near-field configuration), and one where the camera is moved further away (far field configuration). Due to the astigmatism introduced by this set-up, a specific empirical depth correction was necessary to perform quantitative measurements. Considering the optical transfer formalism, it is possible to quantify theoretically the influence of the set-up: the cylindrical geometry induces indeed a difference of the optical transfer matrices from the bubble to the sensor for both transverse x- and y-axes respectively. The matrix for the x-axis is actually given by the following product :

$$\begin{bmatrix}1 & z_4\\0 & 1\end{bmatrix}\begin{bmatrix}1 & 0\\-\frac{1}{f} & 1\end{bmatrix}\begin{bmatrix}1 & z_3\\0 & 1\end{bmatrix}\begin{bmatrix}1 & \overline{H_L'O_L}\\0 & 1\end{bmatrix}\begin{bmatrix}1 & 0\\-\frac{1}{f_L} & 1\end{bmatrix}\begin{bmatrix}1 & \overline{O_LH_L}\\0 & 1\end{bmatrix}$$

$$\times\begin{bmatrix}1 & z_2\\0 & 1\end{bmatrix}\begin{bmatrix}1 & 0\\\frac{n-1}{R_2} & 1\end{bmatrix}\begin{bmatrix}1 & \frac{e}{n}\\0 & 1\end{bmatrix}\begin{bmatrix}1 & 0\\\frac{n_w-n}{R_1} & 1\end{bmatrix}\begin{bmatrix}1 & \frac{z_1}{n_w}\\0 & 1\end{bmatrix} \quad (19)$$

$z_1$ is the distance from the bubble to the first cylindrical diopter (inner face of the pipe). $n_w$ is the index of water, and $n$ the index of the transparent pipe. $R_1$ is the inner radius of the pipe. $e$ is the width of the pipe, $R_2 = R_1 + e$ is the outer radius of the pipe. $z_2$ is the distance from the outer face of the pipe to the main lens of the plenoptic camera. Note that the main lens (objective) is described using a thick system model (with its principal points $H_L'$, $H_L$ around a central position $O_L$ ). $z_3$ is the distance between $O_L$ and the plane of the microlenses. $z_4$ is the distance from the microlens to the sensor. $f_L$ et $f$ are the focus lengths of the main lens and of the microlens respectively. In the case of a multi-focus plenoptic camera (with three sets of microlenses of focus lengths $f_1$, $f_2$, and $f_3$), we obtain three matrices replacing $f$ by these three focus lengths. The optical transfer matrix for the y-axis is given by a similar product except that parameters $R_1$ and $R_2$ are then infinite. The parameters describing the pipe are those of reference [10] : $n_w = 1.33$; $n = 1.51$ ; $R_{1x} = -95mm$ ; $e = 5mm$ ; $R_{2x} = -100mm$. The distance between the pipe and center $O_L$ of the main lens of the plenoptic camera is first $z_2 = 24.3cm$ (also called far-field configuration in reference [10]). For the main lens, we consider $f_L = 100\ mm$, $\overline{O_LH_L} = 40\ mm$ and $\overline{O_LH_L'} = -2\ mm$. The parameters of the multifocus plenoptic camera are : $B_L = 124\ mm$ ; $B = 3\ mm$ ; $\mu = 0.2\ mm$ ; the three microlenses focus lengths $f_1 = 1.56\,B$ , $f_2 = 1.32\,B$ , $f_3 = 1.18\,B$ and a pixel size $p = 5.5\ \mu m$. We have $z_3 = B_L - B$ and $z_4 = B$ (see figure 1 for the definition of these parameters).

Figure 8 shows the resolution of the camera predicted versus the longitudinal position of the initial point object for both x- and y-axes. In each case, three curves are plotted corresponding to the three types of microlenses that constitute the multi-focus camera. The three focus lengths are such that the micro lens' depths of fields touch and eventually slightly overlap in this case. The lines plotted in plain line correspond to the microlens $f_1$, while the dashed lines correspond to the second type of microlenses ($f_2$), and the dash-dotted lines to the third type ($f_3$). We observe clearly the influence of the cylindrical dioptre as the depth of focus does not coincide for the x and y axes. Experimentally, the calibration step performed in reference [10] with a reference target located at the center of the cylinder showed such results: when well-resolved horizontal lines (aligned with the x axis) of the reference target were estimated closer to the camera, blurred vertical lines were estimated further although both types of lines of the target were at the same position.

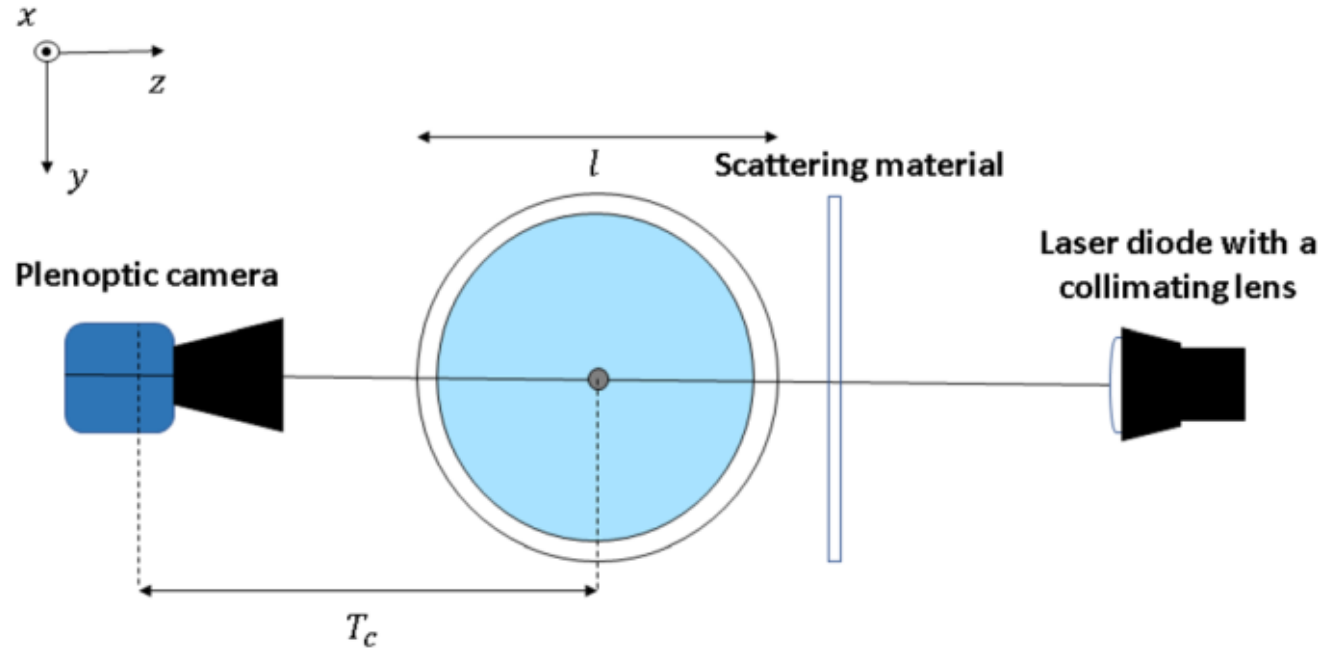

Figure 7: Experimental set-up of reference (top view) [10].

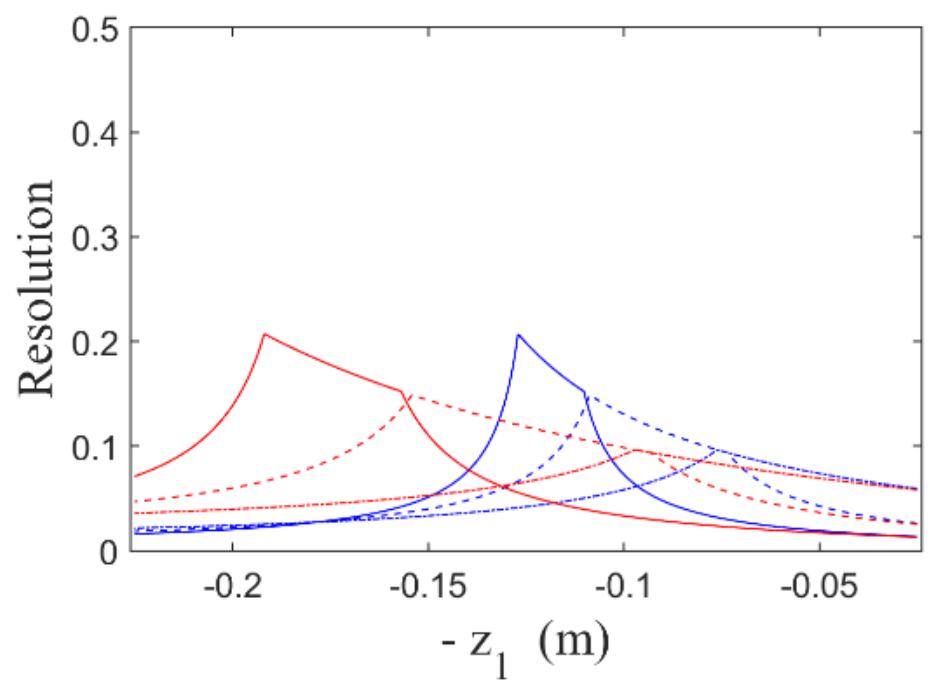

Figure 8: Effective resolution versus the initial object's position in the case of a multifocus 2.0 plenoptic camera. The imaging system is cylindrical. X-axis results in blue. Y-axis results in red.

Let us now compare experimental data to theoretical predictions. The calibration procedure used in reference [10] gives a possible way of quantitative comparisons of disparity and image sizes, as expressed theoretically in relations (14) and (18) respectively. A grid composed of vertical and horizontal lines has been located at different positions z in the cylindrical pipe and the raw images (before any image processing of the data) have been recorded in all cases. One example of experimental raw image obtained with the Raytrix camera is presented in figure 9 for one longitudinal position of the grid in the cylinder. The circles correspond to the images given by the type 1 microlenses of the multifocus microlens array (three types of microlenses are actually present on this array). We could select as well the images given by the type 2 and type 3 microlenses. The different microimages show horizontal and/or vertical lines, depending on the position of each microlens with respect with the grid. These lines are the images of the vertical and horizontal lines of the grid.

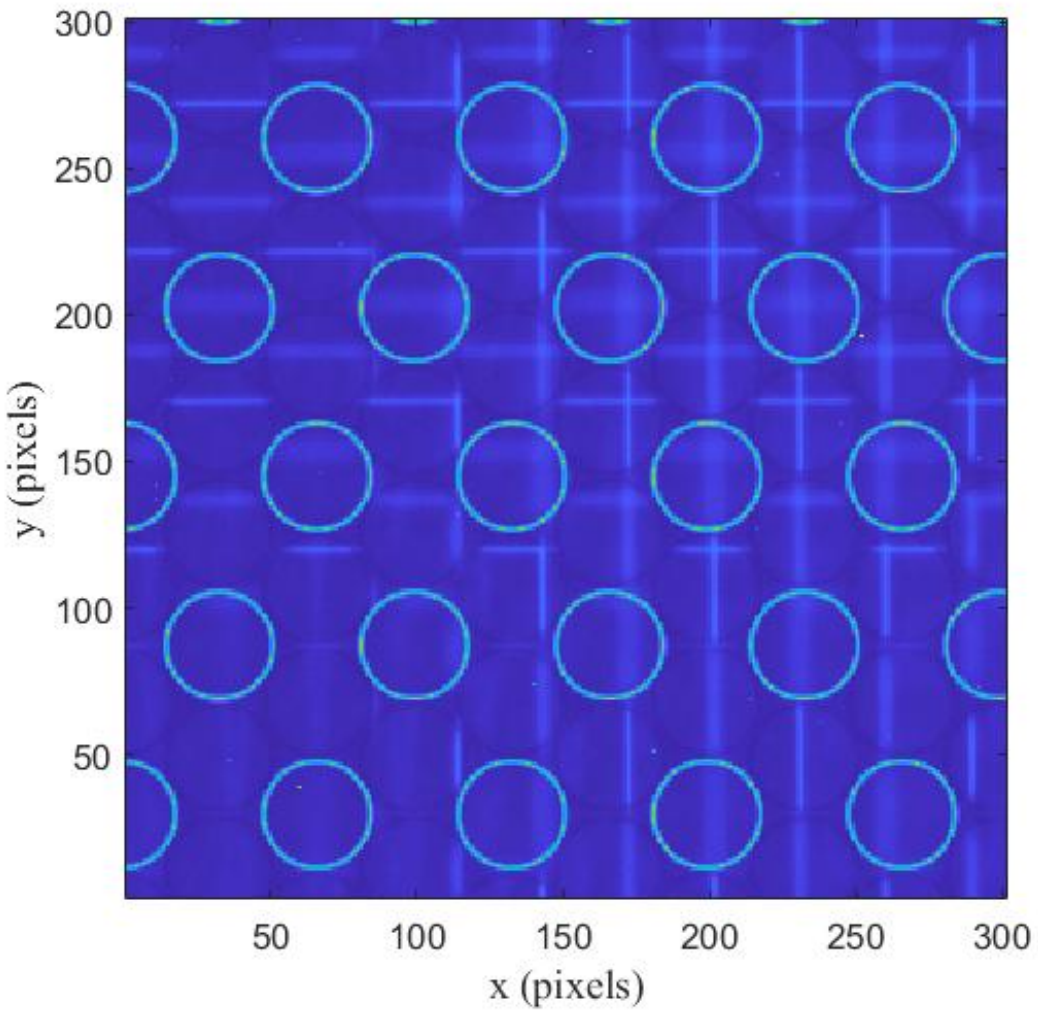

Figure 9: Raw image of a grid in the cylinder at position $z_1$=0.1m.

It is first possible to compare the relative transverse positions of vertical and horizontal lines given by adjacent microlenses. It is thus possible to determine experimentally the disparity for both transverse imaging axes. The difference of the relative positions of vertical lines between adjacent microlenses sharing the same y-coordinate will give the disparity for the x-axis of the optical system. The difference of the relative positions of horizontal lines between adjacent (along the y-axis direction) microlenses will give the disparity for the y-axis of the optical system.

In addition, it is possible to measure the width of each horizontal and vertical imaged line, for the three types of microlenses. The width of the vertical imaged lines is related to the x-axis geometry of the optical system. The width of the horizontal imaged lines is related to the y-axis geometry of the system. Note a difference that will limit the comparison between simulations and experiments : In simulations, the original horizontal and vertical lines of the grid are assumed to be infinitely thin (relation (18)). Experimentally, the lines of the grid have a finite width.

These operations can be repeated for all longitudinal positions of the grid in the cylinder. In addition, the calibration procedure has been done in reference [10] for two positions of the Raytrix plenoptic camera: far-field and near-field experiments. The set of parameters used previously correspond to the experimental far-field configuration. In summary, the disparity for both transverse axes can be determined experimentally versus the position of the grid and compared to values calculated theoretically using relation (14) for both transverse imaging axes. The widths of the imaged lines can be determined experimentally and compared to values calculated theoretically using relation (18) for both transverse imaging axes. The comparison between experiment and theory for the disparity and the line width versus the object position is done for the two imaging configurations of Ref. [10] (near-field and far-field), and reported in figures 10, 11, 13 and 14 : in all figures, theoretical simulations appear in plain line, while the experimental data are plotted as stars. Figures 10-11 present the case of the far field configuration while figures 13 and 14 present the near-field configuration.

The figure 10 shows the disparity versus the position of the grid in the cylinder for both x and y transverse axes, in the case of the far field configuration. Blue lines and marks correspond to the x-axis, while the red ones correspond to the y-axis. Position $z_1$=0.1m corresponds to a grid in the center of the cylinder while position $z_1$=0.07m corresponds to a grid closer to the plenoptic camera. Experimental results are well corroborated by theoretical relations for both axes. Figure 11 shows the width of the imaged lines (Y-axis results in (a) and X-axis results in (b). The lowest value expected theoretically ( $s_0$ ) is never reached experimentally : experimental values measured are always higher than $s_0$. This is attributed to the fact that the object (horizontal or vertical line of the grid) is not perfectly punctual, and to a significant amount of spherical aberration introduced by the cylindrical pipe. Nevertheless, the tendencies of experimental and theoretical curves are in good concordance. For the grid's position $z_1$=17.5 cm in Fig. 11(b), the width of the lines exceeds the diameter of the microlenses (for the three types of microlenses). This is illustrated in figure 12 that shows the raw image of the grid in this case: while the width of horizontal lines is still measurable, the width of the vertical lines is higher than the diameter of the microlenses. So, for the grid's positions $z_1$ > 15 cm in Fig. 11(b), the black star, red cross and blue circle are reported at ordinate 10 to appear in the figure. The exact values are actually higher, but undeterminate. The simulations show a rapidly increasing behavior as well in this domain (when $z_1$ increases).

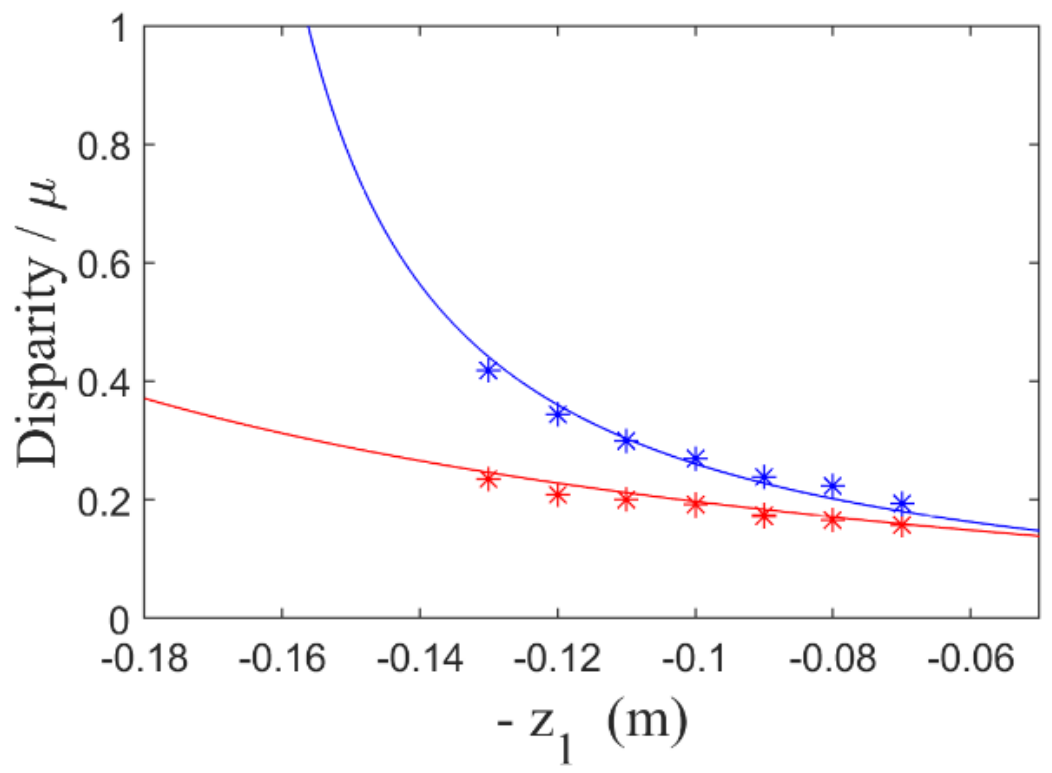

Figure 10: Far-field configuration : disparity versus the initial object's position in the case of a multifocus 2.0 plenoptic camera. The imaging system is cylindrical. X-axis results are in blue, Y-axis results are in red. Plain curves are the theoretical curves.

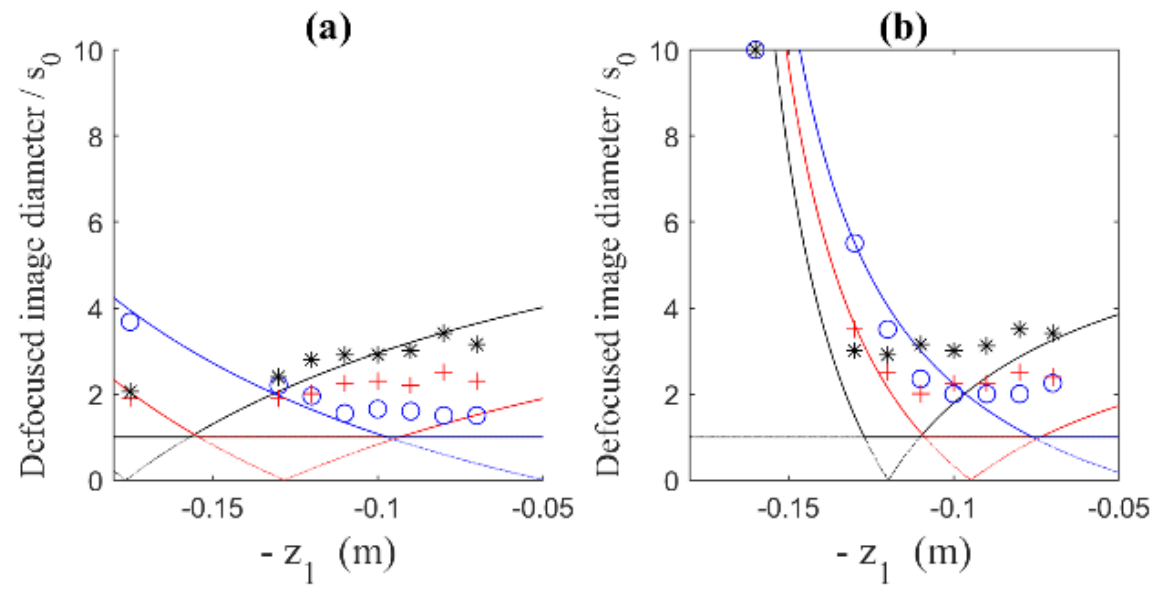

Figure 11: Far-field configuration: image's size of a point object versus the initial object's position in the case of a multifocus 2.0 plenoptic camera. The imaging system is cylindrical. Y-axis results on (a); X-axis results on (b). Black, red and blue correspond to the three types of microlenses.

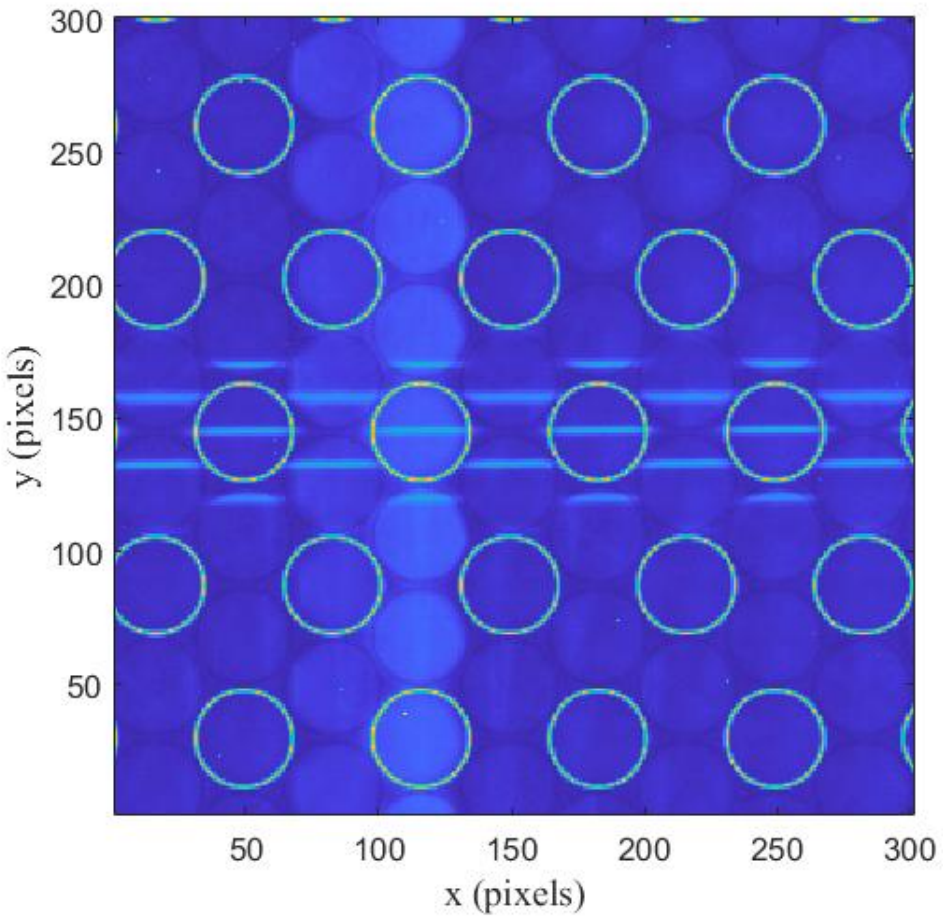

Figure 12: Raw image of a grid in the cylinder at position $z_1$=0.175m.

The parameters used for the near field configuration are the same except: $z_2 = 11cm$ and $B_L = 160\,mm$ (different position and focus adjustment of the camera). Figure 13 shows the disparity versus the position of the grid in the cylinder for both transverse axes x and y of the cylindrical set-up. Experimental values and theoretical predictions are again in good concordance. For the X-axis, only a few positions of the grid lead to a measurable value of disparity between neighbored microlenses of same nature. It is then undetermined (because too high) for higher values of $z_1$. Figure 14 shows the width of the imaged lines (Y-axis results in (a) and X-axis results in (b)). Experimental values appear as black stars, red crosses and blue circles while theoretical values are the plain black, red and blue curves. The three colors on this plot correspond to the three types of microlenses. Experimental values and theoretical predictions show results with very similar tendencies, as observed in the case of the far-field configuration.

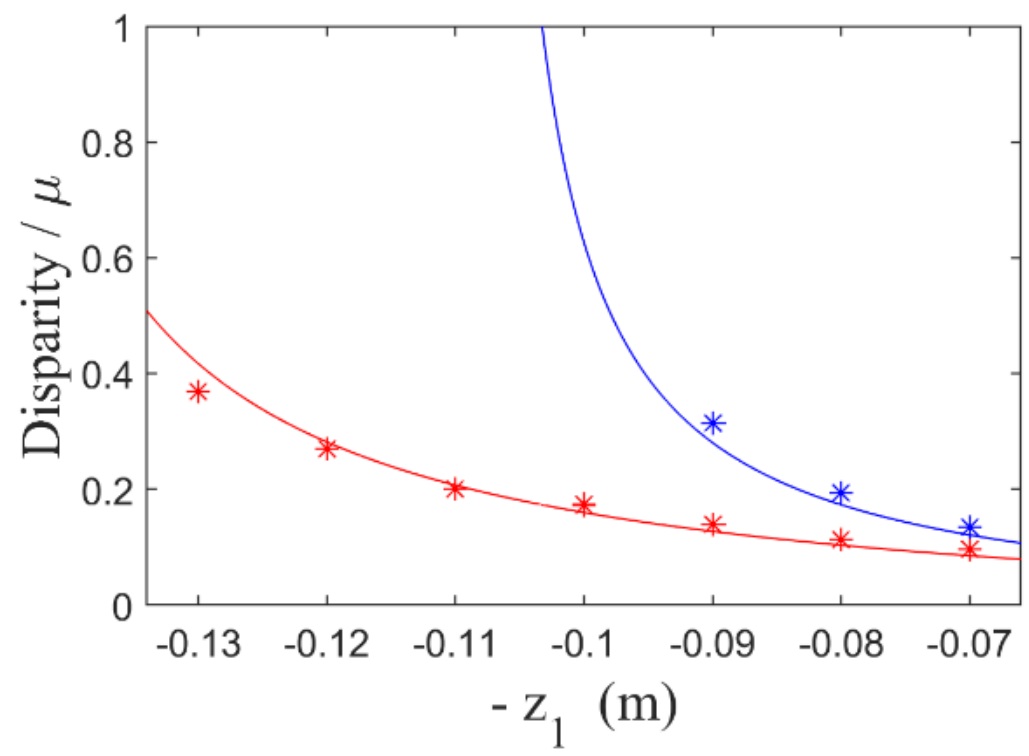

Figure 13: Near-field configuration: disparity versus the initial object's position in the case of a multifocus 2.0 plenoptic camera. The imaging system is cylindrical. X-axis results are in blue, Y-axis results are in red. Plain curves are the theoretical curves.

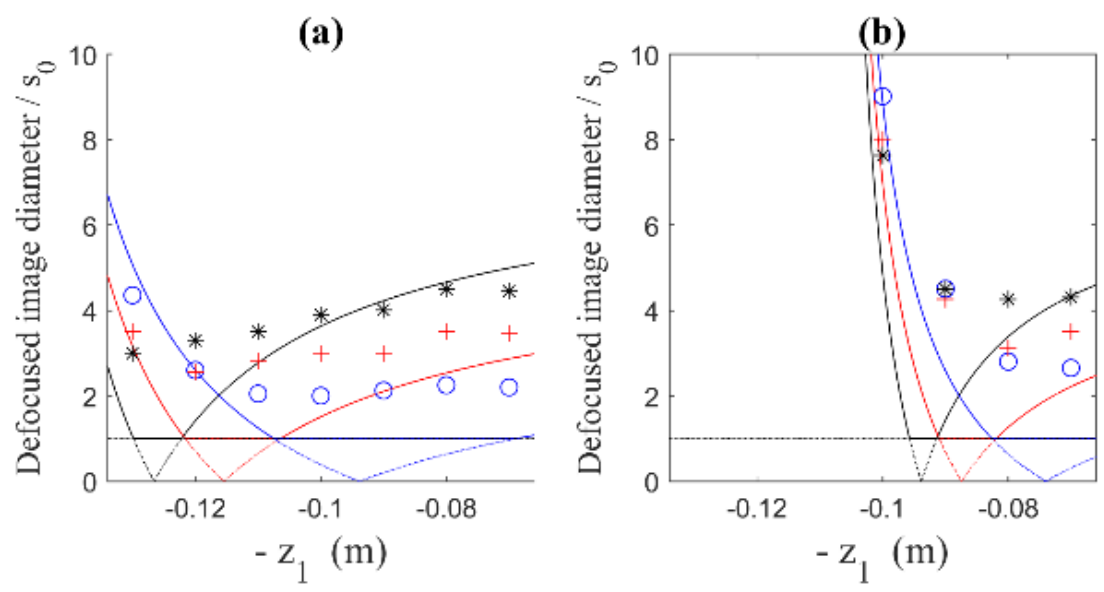

Figure 14: Near-field configuration: image's size of a point object versus the initial object's position in the case of a multifocus 2.0 plenoptic camera. The imaging system is cylindrical. Y-axis results on (a); X-axis results on (b). Black, red and blue correspond to the three types of microlenses.

## 3. CONCLUSION

Light-field imaging is a powerful technique for achieving precise 3D characterization using only a single camera. Across its many application domains (biology, materials science, healthcare, energy, etc.), experimental conditions may require the development of complex imaging systems capable of probing regions of interest that are difficult to access (e.g., combustion or underwater processes), evolve rapidly (such as fluid flows), or span over spatial scales ranging from millimeters to several tens of centimeters. In such contexts, the plenoptic camera often represents just the final component of a broader optical chain, which may include lenses, viewports, relay mirrors, transparent conduits, and media with varying optical properties. Under these conditions, designing the setup and selecting appropriate camera specifications can be particularly challenging. In this study, we employ the optical transfer matrix formalism to derive analytical relationships that directly link the main characteristics of a plenoptic system to the various elements of the overall optical system. Although these relationships do not replace the need for rigorous calibration, they provide a valuable framework for system design and analysis in complex imaging configurations. As such, they are expected to facilitate the broader adoption of light field imaging in increasingly demanding experimental environments.

## Appendix A: depth of focus

The depth of focus is limited by two positions given by parameters $b^+$ and $b^-$ for the final image $X_1$ (resp. $a^+$ and $a^-$ for the intermediate image $X_0$). They correspond to the limit case where the size of the defocused image on the CCD sensor is exactly equal to parameter $s_0$ (see figures A1a and A1b). Geometrically, these positions are given by the following formula :

$$b^+ = B\left[1 - \frac{s_0}{D}\right]^{-1} \quad \text{(A1)}$$

$$b^- = B\left[1 + \frac{s_0}{D}\right]^{-1} \quad \text{(A2)}$$

$$a^+ = B\left[1 - \frac{B}{f} - \frac{s_0}{D}\right]^{-1} \quad \text{(A3)}$$

$$a^- = B\left[1 - \frac{B}{f} + \frac{s_0}{D}\right]^{-1} \quad \text{(A4)}$$

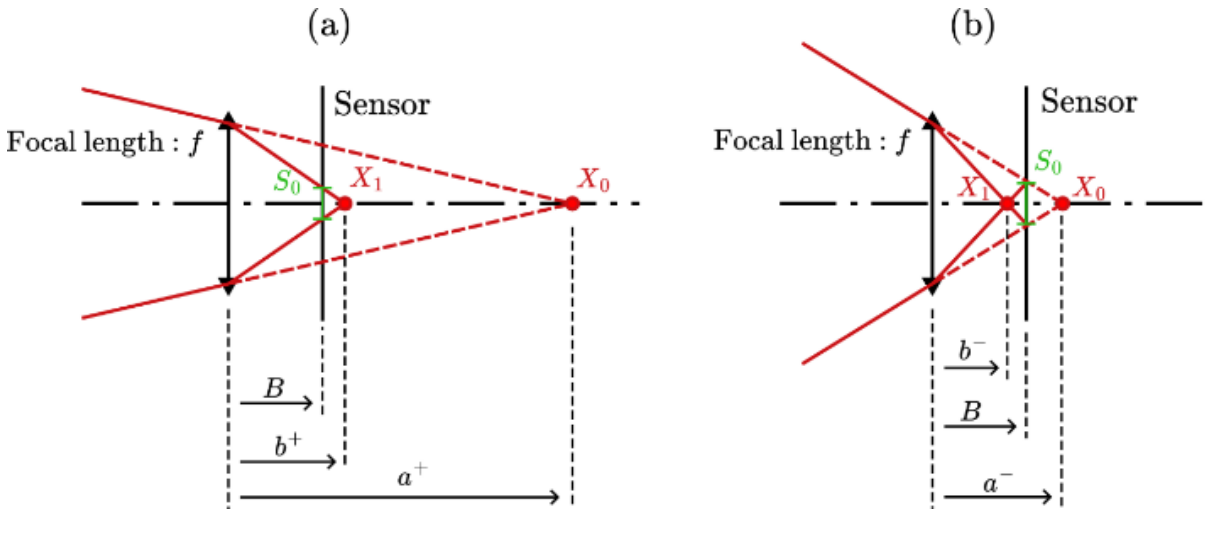

Fig. A1

## Appendix B: Optical transfer matrix formalism

Let us consider a ray that is incident on an optical element at height $y_1$ with angle $\theta_1$. The height $y_2$ and angle $\theta_2$ of the output ray can be calculated in paraxial conditions through:

$$\begin{bmatrix} y_2 \\ n_2\,\theta_2 \end{bmatrix} = \begin{bmatrix} T_{11} & T_{12} \\ T_{21} & T_{22} \end{bmatrix} \begin{bmatrix} y_1 \\ n_1\,\theta_1 \end{bmatrix} \quad \text{(B1)}$$

where $n_1$ and $n_2$ are the indices of the media before and after the optical element respectively, and $\begin{bmatrix} T_{11} & T_{12} \\ T_{21} & T_{22} \end{bmatrix}$ is the optical transfer matrix of the element. If the imaging system is composed of different elements, the optical transfer matrix is the multiplication of the matrices of all elements. With this formalism, we can define independently the optical transfer matrices for the two transverse axes of an optical system.

**Funding.** Carnot ESP MILGIC project.

**Acknowledgments.** Not applicable.

**Disclosures.** Not applicable.

**Data availability.** Data will be made available upon reasonable request.

**Supplemental document**. Not applicable.